%% file: conference_101719.tex
\documentclass[conference, 9pt]{IEEEtran}
\IEEEoverridecommandlockouts
\usepackage{amsmath,epsfig}
\usepackage{caption}
\usepackage{subcaption}
\usepackage{tikz}
\usepackage{booktabs}
\usepackage{xfrac}
\usepackage{url}
\usepackage{graphicx}
\usepackage{cite}
\usepackage{anyfontsize}
\usepackage{amssymb,amsfonts}
\usepackage{pdfpages}
\usepackage{tabularx}
\usepackage{siunitx}
\usetikzlibrary{spy}
\usepackage{nicematrix}
\usepackage{ragged2e}
\usepackage{float}
\usepackage{stfloats}
\usepackage{algorithmicx}
\usepackage{algorithm}
\usepackage[normalem]{ulem}
\usepackage[inkscapeformat=png]{svg}
\usepackage{relsize}
\usepackage{adjustbox}
\usepackage{textcomp}
\usepackage{enumerate}
\usepackage{xcolor}
\usepackage[noend]{algpseudocode}
\usepackage{arydshln}
\usepackage{multirow}
\def\BibTeX{{\rm B\kern-.05em{\sc i\kern-.025em b}\kern-.08em
    T\kern-.1667em\lower.7ex\hbox{E}\kern-.125emX}}
\usepackage{hyperref}
\begin{document}
\title{Lightweight Video Denoising Using a Classic Bayesian Backbone
\thanks{This research is supported by Science Foundation Ireland in the \href{www.adaptcentre.ie}{ADAPT Centre} (Grant 13/RC/2106) at Trinity College Dublin.}
}

\author{\IEEEauthorblockN{1\textsuperscript{st} Cl\'ement Bled}
\IEEEauthorblockA{\textit{dept. of Electrical and Electronic Engineering} \\
\textit{Sigmedia}\\
Trinity College, Dublin, Ireland \\
bledc@tcd.ie}
\and
\IEEEauthorblockN{2\textsuperscript{nd} Fran\c{c}ois Piti\'e}
\IEEEauthorblockA{\textit{dept. of Electrical and Electronic Engineering} \\
\textit{Sigmedia}\\
Trinity College, Dublin, Ireland \\
pitief@tcd.ie}
}

\maketitle

\begin{abstract}
In recent years, state-of-the-art image and video denoising networks have become increasingly large, requiring millions of trainable parameters to achieve best-in-class performance. Improved denoising quality has come at the cost of denoising speed, where modern transformer networks are far slower to run than smaller denoising networks such as FastDVDnet and classic Bayesian denoisers such as the Wiener filter.

In this paper, we implement a hybrid Wiener filter which leverages small ancillary networks to increase the original denoiser performance, while retaining fast denoising speeds. These networks are used to refine the Wiener coring estimate, optimise windowing functions and estimate the unknown noise profile. Using these methods, we outperform several popular denoisers and remain within 0.2 dB, on average, of the popular VRT transformer. Our method was found to be over x10 faster than the transformer method, with a far lower parameter cost. 
\end{abstract}

\begin{IEEEkeywords}
Video Denoising, Image Sequence Denoising, Wiener Filter
\end{IEEEkeywords}

\section{Introduction}
\label{sec:intro}

\input{2_intro}



\section{Background}
\input{3_background}


\section{Enhanced Wiener Denoising for Video}
\label{sec:method}
\input{4_method}

\section{Experiments/Results}
\label{sec:results}
\input{5_experiments_results}

\label{sec:conclusion}


\bibliographystyle{IEEEtran}
\clearpage
\bibliography{refs}
\end{document}

%% file: 2_intro.tex
Denoising remains a crucial step in many applications of image and video processing, 
from the smartphone camera ISP pipeline to denoising tools of the post-production industry. More recently, the mass adoption of over-the-top streaming services such as
Netflix and Disney+, as well as social media driven by user-generated content such
as YouTube, Twitch.tv, Instagram and Facebook have placed greater importance on efficient video
encoding, where denoising is essential in reducing frame entropy and reducing the bandwidth necessary to distribute and receive content.

Classic denoisers which rely on Bayesian modelling and frequency filtering such
as Wiener
filters~\cite{pratt1972generalized,king1983wiener,giger1984investigation,benesty2010study} and
Wavelet filters~\cite{mallat1989theory,combettes2004wavelet,malfait1997wavelet},
or those which use patch similarity, as in BM3D~\cite{dabov2007image},
V-BM4D~\cite{maggioni2012video} and
VNLB\cite{arias2018video}, have recently been outperformed by deep learning
approaches~\cite{8803314,tassano2019dvdnet,tassano2020fastdvdnet,claus2019videnn,yue2020supervised,vaksman2021patch,liang2022vrt,yue2023rvideformer,wang2019edvr}. In
2019, Maggioni et al. put forward DVDNet~\cite{tassano2019dvdnet}, which
outperformed VNLB\cite{arias2018video} using a two-step CNN architecture: a
spatial denoising network applied to motion-compensated frames, followed by a
temporal denoising step which consolidates the output of three spatially
denoised adjacent frames into a single frame. Originally 1.3M parameters in
total, FastDVDNet~\cite{tassano2020fastdvdnet} increased the network size to
2.5M in total, opting for a U-Net architecture in its denoising blocks and
replacing motion compensation with overlapping, multi-frame input blocks.

Inspired by DVDNet, similar networks such as Videnn~\cite{claus2019videnn} (3.5M
parameters) and PaCNet~\cite{vaksman2021patch} (2.9M parameters) have been
proposed. More recently, following the success of image vision
transformers such as SwinIR~\cite{liang2021swinir} (Liang et al.) and Restormer~\cite{zamir2022restormer} (Zamir et al.), Liang et al. put forward the Video Restoration
Transformer\cite{liang2022vrt} (VRT), achieving best-in-class results with a
network of 35.6M parameters. Unlike image denoisers, the most popular video denoising algorithms (VRT, DVDNet, FastDVDNet, VNLB) are non-blind, meaning the user is required to supply the denoiser with a measure of the noise variance.

While transformer networks achieve greater PSNR quality scores, they are slower to run than smaller networks (See Table~\ref{tab:summary_table}), and their increased parameter count results in high video memory consumption when running inference on high-resolution images,  limiting the hardware on
which they may be deployed. 


In recent work from Bled and Piti\'e~\cite{10222826}, it was demonstrated that this trend of increasingly larger networks is not a fatality and that the original Wiener filter can actually be optimised to achieve 
performances close to popular image denoising DNNs such as
DnCNN~\cite{zhang2020deep}.

In this paper, we adopt a similar approach for video denoising and explore how the Wiener filter could be used as the backbone of a state-of-the-art video denoiser architecture. We reconsider all tuneable parameters of Bled's Wiener filter, taking special care to optimise for denoising speed as temporal data is introduced. We introduce trainable window functions, 4D FFTs and 3D CNNs, as well as an ablation study on the use of motion compensation in video denoisers. We also modify Bled's blind denoiser to generalise to Video denoising.

Our key contribution is the implementation of a denoiser which demands far fewer parameters (0.29 M) than current denoising networks, outperforming DVDNet, FastDVDNet, and VNLB, on average in terms of PSNR. We outperform all tested networks in SSIM and achieve greater performance than the Vision transformer~\cite{liang2022vrt} at high noise levels. 



%% file: 3_background.tex
\subsection{Baseline Video Wiener Filter}
\label{sec:background}

Given a noisy signal $y$, composed of the original, unknown signal $x$, and
additive noise $n$, $y = x + n$; the Wiener filter
\cite{wiener1949extrapolation} defines a linear, minimum mean square error
(MMSE) optimal filter. Assuming that the image sequence and noise signal are
second-order stationary and decorrelated, the optimal IIR Wiener filter is given
by the following transfer function $H(\omega_1,\omega_2,\omega_t)$:
\vspace{-0.5em}
\begin{equation}
    H(\omega_1,\omega_2,\omega_t) = \frac{ P_{xx}(\omega_1,\omega_2,\omega_t) }{ P_{yy}(\omega_1,\omega_2,\omega_t)}\,,
\end{equation}
where $P_{yy}$ and $P_{xx}$ are the power spectrum densities at spatial and
temporal frequencies $\omega_1,\omega_2,\omega_t$ at frame $t$ for the input
signal $y$ and original signal $x$.  In practice, the PSD of the unknown, clean
signal is estimated with the following \textit{coring} function at each frequency $(\omega_1, \omega_2, \omega_{t})$:
\vspace{-0.1em}
\begin{equation}\label{eqn: wiener_coring}
    \hat{P}_{xx} = \max( P_{yy} - P_{nn}, 0).
\end{equation}
The noise PSD $P_{nn}$ can be measured offline, but if it is Additive White
Gaussian (AWG), the PSD is a constant $P_{nn} \propto \sigma^2 $, where $\sigma$
is the noise standard deviation (STD).

The use of Wiener filter filter for video denoising was first popularised by
Kokaram~\cite{kokaram19943d}, which made use of motion compensation algorithms
as a preprocessing measure. We summarise this implementation in
Alg.~\ref{alg:WienerBaseline} (the symbols $\odot$ and $\oslash$ denote
element-wise multiplication and divisions in the blocks). As image sequences are
not stationary processes, the sequence must be broken into blocks
(\textit{eg.} 32\texttimes32) to approximate a stationary signal. An analysis window is used for the frequency analysis of the block. All processed blocks are overlapped and added, using a spatial interpolation windowing function called the synthesis window. Kokaram used the same half-cosine for the synthesis and frequency analysis window, as it allows for some simplification in the overlap-add step as the weights sum up to 1.

{ \setlength{\textfloatsep}{5pt}
\begin{algorithm}[t]
\small 
\caption{Kokaram's Video 3D Wiener Filter\cite{kokaram19943d}\label{alg:WienerBaseline}
}
\begin{algorithmic}[1]
    \Require Noisy image seq, noise STD $\sigma$, Block Size
    \State $w(t,h,k) \gets RaisedCosine(t,h,k)$ \Comment{windowing definition}
    \ForAll {frames in seq}
    \State $y \gets $ 3D framebuffer made of current grayscale frame and 4 nearby motion
    compensated neighbouring frames 
    \ForAll {blocks ${\bf y}$ in $y$, for stride=BlockSize/2}
    \State $\bar{y} \gets \mathrm{mean}({\bf y})$ \Comment{predicts block mean}
    \State ${\bf y}_w \gets ({\bf y} - \bar{y}) \odot {\bf w}$ \Comment{windowing}
    \State ${\bf Y} \gets FFT3D({\bf y}_{w})$ 
    \State ${\bf P}_{yy} \gets {\bf Y} \odot {\bf Y}^*$
    \State ${\bf P}_{nn} \gets \hat{\sigma}^2 \| {\bf w} \|^2$
    \State ${\bf P}_{xx} \gets \max({\bf P}_{yy} - P_{nn}, 0)$ \Comment{coring}
    \State $\hat{\bf x}_{w} \gets iFFT3D({\bf Y} \odot {\bf P}_{xx} \oslash {\bf P}_{yy}) + \bar{y}  {\bf w}$
    \State ${\hat x} \gets \mathrm{overlap\_add}({\bf w} \odot \hat{\bf x}_{w})$
    \Comment{combine blocks}
    \EndFor    
    \EndFor
\end{algorithmic}
\end{algorithm}
}

\subsection{Improving the Wiener Baseline}

Recently Bled and Piti\'e~\cite{10222826} demonstrated that this baseline Wiener filter for
\textit{image-denoising} could be improved by about +2.8dB PSNR by making a
number of small adjustments. These include directly processing R, G, and B channels in a separate dimension, taking denser block overlaps with a
quarter block stride instead of the typical half-block stride, using a Gaussian
analysis and interpolation windows in place of the half-cosine windows, using
median estimation over pixel averaging for DC-offset removal before the FFT
transform and, lastly, apply the filter at different scales.

They also outline that a further +0.5dB could be obtained by refining the estimated
Wiener coring kernel $H(\omega_1,\omega_2)$ with a very small convolutional
network, thus bringing the overall performance of the image denoiser on par with
popular networks such as DnCNN~\cite{zhang2017beyond}, but with fewer
network parameters.

\begin{figure}[t]
  \includegraphics[width=\linewidth]{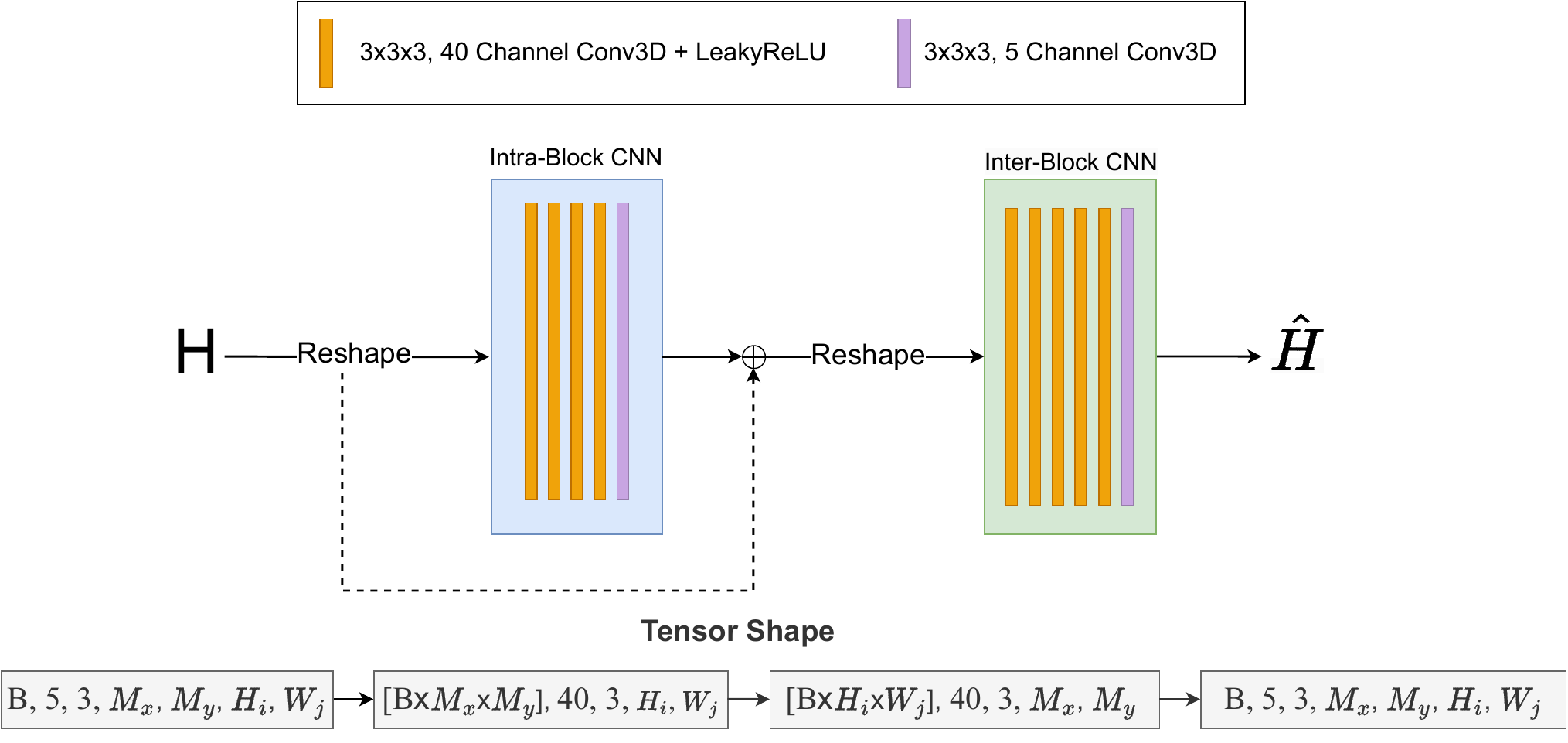}
  \caption{The two-stage coring refinement network architecture used to optimise the initial prediction of the coring function $H(\omega_1, \omega_2)$. }
  \label{fig:coring_cnn}
\end{figure}

{ \setlength{\textfloatsep}{5pt}
\begin{algorithm}[t]
\small 
\caption{Our Video 4D Wiener Filter\label{alg:Wiener4D}
}
\begin{algorithmic}[1]
    \Require Noisy image seq, noise STD $\sigma$, Block Size
    \State $w_a(h,k,t) \gets \exp\left( - \alpha_a(h^2+k^2) \right)$ \Comment{analysis window}
    \State $w_s(h,k,t) \gets \exp\left( - \alpha_s (h^2+k^2) \right)$ \Comment{synthesis window}    
    \ForAll {frames in seq}
    \State $y \gets $ 4D framebuffer made of current RGB frame and 4 motion
    compensated neighbouring RGB frames
    \ForAll {blocks ${\bf y}$ in $y$, for stride=BlockSize/4}
    \State $\bar{y} \gets \mathrm{median}({\bf y})$ \Comment{predicts block' DC offset}
    \State ${\bf y}_w \gets ({\bf y} - \bar{y}) \odot {\bf w}_a$ \Comment{analysis window}
    \State ${\bf Y} \gets \mathrm{FFTn}({\bf y}_{w})$ 
    \State ${\bf P}_{yy} \gets {\bf Y} \odot {\bf Y}^*$
    \State ${\bf P}_{nn} \gets \hat{\sigma}^2 \| {\bf w} \|^2$
    \State ${\bf P}_{xx} \gets \max({\bf P}_{yy} - P_{nn}, 0)$ \Comment{coring}
    \State $\hat{\bf x}_{w} \gets \mathrm{iFFTn}({\bf Y} \odot {\bf P}_{xx} \oslash {\bf P}_{yy}) + \bar{y}$
    \State $x_{\mathrm{all}} \gets \mathrm{overlap\_add}({\bf w}_s \odot \hat{\bf x}_{w})$ \Comment{combine blocks}
    \State $w_{\mathrm{all}} \gets \mathrm{overlap\_add}({\bf w}_s \odot {\bf w}_a)$
    \Comment{combine windows}
    \EndFor
     \State $\hat{x} \gets \frac{x_{\mathrm{all}}}{w_{\mathrm{all}}}$ \Comment{denoised frame}
    \EndFor    
\end{algorithmic}
\end{algorithm}
}

%% file: 4_method.tex

\subsection{A Video Wiener 4D Backbone Network}
\label{sec:method_4d_baselineopt}

In this work, we propose to extend the idea from Bled et al. to form a video denoising network based on a Wiener Filter backbone. As we introduce the temporal dimension, we must revisit the optimisation made by Bled, as previous optimal values no longer apply. We also take extra care to optimise for denoising speed.

We start from the baseline 3D Wiener video denoising filter implementation by Kokaram and include some of the ideas proposed in~\cite{10222826} to form a new method that we will
call Wiener 4D.

As the name suggests, we first expand the Wiener filter to handle colour as an additional dimension, thus Kokaram's 3D FFT becomes a 4D FFT, using the RGB channels as the third dimension and a temporal window of 5 frames as the fourth dimension. While the filter returns five filtered frames, only the target frame is saved. 

A summary of our 4D Wiener Filter is outlined in Alg.~\ref{alg:Wiener4D}. A notable difference with Kokaram's Wiener filter baseline is that our window functions need to be explicitly normalised to one in the synthesis step. This is because we also explore the choice of analysis and synthesis windowing in terms of window overlap stride, window size, and, window shape. In section~\ref{sec:window_opt}, we introduce trainable 3D windows and evaluate their performance compared to Raised-Cosine windows and Gaussian windows. 

In section~\ref{sec:window_opt}, we also show that the method of DC-offset removal, a necessary preprocessing step of the FFT, can have some significant impact. Because of range clipping, noise is biased in the black regions and white regions. This bias is rarely addressed in the literature but it usually means that denoised images blacks are not dark enough. In this paper, we show that using the median for DC-offset estimation is surprisingly effective in video denoising, suppressing any visible bias, and leading to similar performance as when using the Ground-Truth DC values.

\subsection{Video Wiener Coring Refinement Network}\label{subsec:coring_opt}

As an alternative to the default Wiener coring function of Eq.~(\ref{eqn: wiener_coring}), we propose, as in~\cite{10222826} a lightweight coring post-processing network that operates on the 4D spectral tensor. This network aims to reduce potential ringing artefacts caused by the default coring estimation errors. The network takes in the MSE-optimal Wiener filter transfer function $H(\omega_1, \omega_2, \omega_t, \omega_c)$ as computed by Eq.~(\ref{eqn: wiener_coring}) as a 4D tensor for the spatial frequencies $\omega_1,\omega_2$, temporal frequency $\omega_t$ and RGB channel frequency $\omega_c$, and predicts a new estimate, $\hat{H}$.  

A simplified block diagram of our two-stage network is shown in Figure~\ref{fig:coring_cnn}. 
The network architecture significantly differs from~\cite{10222826} because we have now to deal with the temporal dimension. The network accepts a Wiener tensor $H$, of shape [\textit{B\texttimes T\texttimes C\texttimes
    M\textsubscript{x}\texttimes M\textsubscript{y}\texttimes H\texttimes W}], which
is rearranged to consolidate the \textit{M}\textsubscript{x}\texttimes \textit{M}\textsubscript{y}
overlapping analysis windows, to the batch dimension, \textit{B}, to create a tensor of
shape [(\textit{B}\texttimes\textit{M}\textsubscript{x}\texttimes \textit{M}\textsubscript{y})\texttimes\textit{T}\texttimes\textit{C}\texttimes\textit{H}\texttimes\textit{W}]. This allows us to refine the filter via 3D trainable convolutions. In the second stage of refinement, the tensor is rearranged to have shape [\textit{B\textsubscript{HW}\texttimes T\texttimes C\texttimes M\textsubscript{x}\texttimes M\textsubscript{y}}] such that we may refine the network via inter block pixel relationships. We name two parts of the network the \textit{intra-block} and the \textit{inter-block} stages respectively.

The network is composed of 11, 3D-Convolution layers, each paired with a LeakyReLU activation, with the exception of the last layer in each block, from which they are omitted. 40 filters/channels are used throughout and each convolution is bias-free to improve generalisation
on unseen data~\cite{mohan2020robust}. We train the network using the weighted sum of two L1 losses: firstly that of the target centre frame, and secondly, that of the entire 5-frame sequence. 
For both denoising stages, the final network sums to (139,320+139,995) 279,315 parameters.

\subsection{Blind denoising}
\label{subsec:blind_denosier}
As is still typical in denoisers today, the classic Wiener filter is a `nonblind' filter, requiring an estimate of the degraded image noise standard deviation. As denoising networks move towards blind denoising, we also implement a blind Wiener filter, requiring no extra inputs from the user.

A small ancillary 2D CNN, is implemented for this purpose, which takes in the centre target frame of the 5-frame sequence and returns a noise standard deviation map of the same size. The map is repeated for the outer four frames and fed to the Wiener filter. Unlike the user-input noise STD, the predicted noise map is not constrained to a single value across the 3-channel image. The network is trained alongside the coring refinement, with the additional L1 loss between the predicted noise STD map and the ground truth uniform map added. To allow the STD network to be trained unconstrained by the uniform standard deviation loss and to maximise output quality, a second training stage is run with only the target frame loss. All five noise level datasets are combined to train this network.

This ancillary network consists of only four 2D convolution layers and three LeakyReLU activations. The network contributes only 8,280 extra parameters to the coring refinement network, increasing its size to 287,595 parameters.

\subsection{Motion Compensation and Multi-Scale Averaging}
\label{sec:mocomp_and_multiscale}
Many video denoising networks still implement forms of motion estimation to map pixels in non-target frames to their position in the target frames. This spatial alignment is often necessary to increase performance in classic Bayesian filters which rely on temporal consistency but its effect on denoising networks is unclear. To assess the impact of motion compensation in trained networks, we compare the denoising performance of our optimised Wiener filter with, and, without motion compensation, for both our trained Wiener filter and our untrained filter.

Lastly, we examine the performance benefits of a multi-scale Wiener filter, whereby the image is denoised at multiple block sizes, and the outputs are averaged. Capturing multi-scale frequency information in this manner increases the amount of information available to the denoiser and creates a smoother final image.

%% file: 5_experiments_results.tex
In this section, we iterate through the optimisations made to the 4D Wiener filter, measuring quality improvements at each step. We evaluate our denoiser using ten, 64-frame sequences, taken from a combination of Derf's Collection~\cite{montgomery1994xiph} (HD, gaming) and the BVI (SynTex~\cite{ma2021bvi}, DVC~\cite{katsenou2020bvi}) datasets. Each clip is centre-cropped to 500\texttimes500 for evaluation. For training, we choose 173 uncropped, full-length videos from the corpus, omitting the test sequences. Additive Gaussian noise is applied to the datasets at standard deviations of $\sigma$ = [10, 20, 30, 40, 50]. At training time, five frames are randomly selected from each sequence and cropped into batches of 5\texttimes128\texttimes128.  



\subsection{Optimising Block Overlaps and Block Size}
We first optimised the 4D filter for the overlap between denoised temporal blocks. We measure the stride of the sliding analysis window as a fraction of the block size, 32\texttimes32, from 1/2 (Kokaram's standard) a block width to 1/8 of a block. Our quality measurements are taken as the average across the 10 sequence dataset at a noise STD=20. 

In Table~\ref{tab:overlap_win} we observe a +0.17 dB PSNR gain by reducing the stride to 1/3 of a block width. Further decreasing the stride provides little improvement in quality while greatly increasing denoising time; a quarter stride increases performance by only 0.01 dB and increases the denoising time by 6.7 seconds, as more analysis blocks must be denoised.
\begin{table}[t]
\centering
\resizebox{0.75\width}{!}
{%
\begin{tabular}{@{}lllr@{}}
\toprule
Stride & PSNR (dB) & SSIM ([0-1]) & Time (s) \\ \midrule
1/2                  & 31.56      & 0.83756     & 5.39                     \\
1/3                  & 31.73      & 0.84316     & 13.26                     \\
1/4                  & 31.74      & 0.84360     & 19.94                     \\
1/5                  & 31.75      & 0.84378     & 35.19                    \\
1/6                  & 31.75      & 0.84383     & 16.83                    \\
1/7                  & 31.75      & 0.84384     & 49.80                    \\
1/8                  & 31.75      & 0.84384     & 78.46                    \\ \bottomrule
\end{tabular}%
}
\caption{Study of Wiener window stride as a fraction of block size versus output quality (PSNR / SSIM). A block size of 32\texttimes32 is used. Quality measurements are taken as an average of the 10-sequence dataset. Time measurements are taken as the sum of denoising time for the 10 sequences.}
\label{tab:overlap_win}
\end{table}

\begin{figure}[t]
    \includegraphics{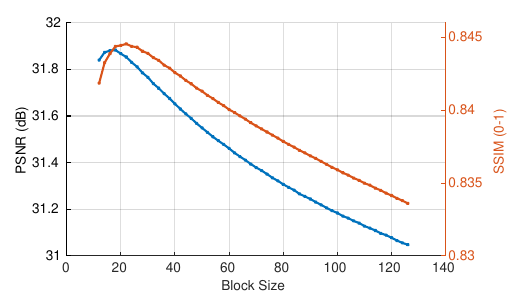}
    \label{fig:subfig_blocksize_vs_qual}

\caption{Graph Measuring Wiener window block size versus output quality in terms of PSNR (dB) and SSIM (0-1). Quality measurements are taken as an average of our 10-sequence test set. $\sfrac{1}{4}$ overlap stride used.}
\label{fig:win_size_figs}
\end{figure}

Next, we optimise for window size using the same noise profile, $\sigma$ = 20. In Fig.\ref{fig:win_size_figs} we plot denoising quality w.r.t block size for both PSNR and SSIM. We observe that PSNR peaks at a window size of 18 (31.88 dB), and decreases as the analysis block size increases to 126 (31.05 dB). This optimal is + 0.14 dB greater than the previous implementation. SSIM peaks at a window of size 22 (0.8445), and the lowest quality (0.8336) is also recorded at the largest window size, 126. Increasing window size was not found to significantly increase or decrease denoising time. While larger windows result in more expensive FFT transforms, fewer windows are required to cover the frames. 
\label{sec:window_opt}


\subsection{Optimising DC Offset Removal and Window Shape}
\label{subsec:result_window_dc}
As mentioned in Section~\ref{sec:window_opt}, it is necessary to zero-mean the temporal block before applying the FFT. For this purpose, we compare classic mean subtraction to median subtraction. To measure the performance lost by taking the noisy mean, we record the output quality when the denoiser is provided with the unseen ground truth mean. 
In table~\ref{tab:tab:DC_offset2}, we show that for all noise levels, using the median of the block increases denoising performance. This effect is most noticeable at higher noise standard deviations, where pixels close to brightness boundaries deviate further from their ground truth values. At a noise level of STD=50, we note a \textbf{+ 0.27 dB} increase in performance over using the noisy mean.
\begin{table}[t]
\centering
\resizebox{0.7\columnwidth}{!}
{%
\begin{tabular}{@{}lllll@{}}
\toprule
      &  \multicolumn{1}{c}{Mean} & \multicolumn{1}{c}{Median} & \multicolumn{1}{c}{GT Mean}  \\ 
STD & PSNR / SSIM    & PSNR / SSIM   & PSNR / SSIM   \\ \midrule
10    &  35.61 / 0.9099 & \textbf{35.63} / \textbf{0.9105} & 35.63 / 0.9105 \\
20    &  31.87 / \textbf{0.8445} & \textbf{31.90} / 0.8391          & 31.90 / 0.8449 \\
30    &  29.55 / 0.8527 & \textbf{29.65} / \textbf{0.8535} & 29.65 / 0.7908 \\
40    &  27.71 / 0.7424 & \textbf{27.90} / 0.7445 & 27.89 / \textbf{0.7760} \\
50    &  26.09 / 0.7007 & \textbf{26.36} / \textbf{0.7040} & 26.35 / 0.7039 \\ \bottomrule
\end{tabular}%
}

\caption{Study of DC offset removal strategies as a preprocessing step to the
  Wiener Filter using a 32\texttimes32 window. For comparison, the ground truth
  mean in the final column uses the unknown clean image to generate the DC offset. Each result is the 10-sequence average quality.}
\label{tab:tab:DC_offset2}
\end{table}

Next, we study the impact of window shape on denoising performance. In their original paper, Kokaram used a raised cosine window which acted as both the analysis and spatial interpolation window for the overlapping blocks. As mentioned in Section~\ref{sec:window_opt}, changing the window stride means that overlapping blocks no longer sum to one. Instead, we use separate analysis and interpolation window pairs, and, to ensure the overlapping windows sum to one, a normalising weight map is applied to the reconstructed frame. 

In addition to the half cosine and Gaussian windows used by Kokaram and Bled respectively, we introduce two new analysis-interpolation window pairs: trainable Gaussian (non-isotropic) windows and trainable isotropic windows. The trainable Gaussian windows are initialised as normal Gaussian windows and their weights are set to be trainable. The trainable isotropic window is initialised as a 1D Gaussian window, with the final window being interpolated onto 2D space. In both cases, the windows are saved post-training and added to the filter as fixed weights. \\

\noindent \textbf{Training:} The weights are trained using the same loss function described in section~\ref{subsec:coring_opt}, with the AdamW optimiser~\cite{loshchilov2018fixing} and a cosine Annealing learning rate scheduler which reduces the learning rate from $1e^{-3}$ to $1e^{-5}$ every 300 epochs, over 1200 epochs. 

In table~\ref{tab:win_shapes} we show that the original Raised Cosine window is outperformed by our trainable Gaussian window by \textbf{+ 0.21 dB}, a small improvement over the non-trained Gaussian window. The isotropic window also outperforms the half-cosine window but we were unable to exactly match Gaussian windowing in out training.  

\begin{table}[t]
\centering
\resizebox{0.9\linewidth}{!}{%
\begin{tabular}{@{}llllll@{}}
\toprule
 & \multicolumn{1}{c}{Cosine} & \multicolumn{1}{c}{Gaussian} & \multicolumn{1}{c}{Trainable} & \multicolumn{1}{c}{Isotropic} \\ 
Scene     & PSNR / SSIM    & PSNR / SSIM    & PSNR / SSIM    & PSNR / SSIM    \\ \midrule
India     & 31.45 / 0.9205 & 31.72 / 0.9246 & 31.82 / 0.9257 & 31.62 / 0.9234 \\
Market    & 32.36 / 0.8524 & 32.33 / 0.8489 & 32.42 / 0.8508 & 32.33 / 0.8508 \\
DOTA2     & 35.28 / 0.9055 & 35.46 / 0.9073 & 35.54 / 0.9076 & 35.31 / 0.9068 \\
Hamster   & 30.69 / 0.5360 & 30.75 / 0.5358 & 30.79 / 0.5359 & 30.62 / 0.5342 \\
Shopping  & 31.63 / 0.9207 & 31.94 / 0.9271 & 32.07 / 0.9269 & 31.83 / 0.9250 \\
Football  & 29.37 / 0.8890 & 29.40 / 0.8891 & 29.47 / 0.8907 & 29.40 / 0.8897 \\
Tree      & 32.85 / 0.8550 & 33.03 / 0.8568 & 33.05 / 0.8570 & 32.97 / 0.8571 \\
Minecraft & 28.34 / 0.7416 & 28.30 / 0.7379 & 28.31 / 0.7384 & 28.28 / 0.7387 \\
Bridge    & 32.25 / 0.9238 & 32.48 / 0.9261 & 32.63 / 0.9268 & 32.41 / 0.9258 \\
Christmas & 33.49 / 0.8955 & 33.63 / 0.8961 & 33.69 / 0.8968 & 33.59 / 0.8967 \\ \midrule
Mean      & 31.77 / 0.8440 & 31.90 / 0.8450 & \textbf{31.98 / 0.8457} & 31.84 / 0.8448 \\ \bottomrule
\end{tabular}%
}
\caption{Windowing function vs. denoised quality (PSNR / SSIM). Trainable+ denotes a trained window constrained to positive values only. We evaluate all results on our test set of STD $\sigma$=20, at a window size of 32\texttimes32 using a quarter overlap.}
\label{tab:win_shapes}
\end{table}
\subsection{Coring Refinement Network and Blind Denoising}
We now evaluate the performance of the coring refinement network, as described in Section~\ref{subsec:coring_opt}. The network is trained using the same scheme as outlined in the previous section (\ref{subsec:result_window_dc}), with a 1/3 block stride and the learned Gaussian windows. The window sizes are set to 16\texttimes16 to maximise performance.

For our non-blind denoiser, we train five networks separately, each on a separate Gaussian noise profile: $\sigma$ = [10, 20, 30, 40, 50]. These denoisers are non-blind and require the user to provide the denoiser with a noise STD.  In Table~\ref{tab:summary_table} we show that WienerNet performance is on average \textbf{+ 3.5 dB} (+ 0.1311 SSIM) greater than the optimised baseline Wiener (non-coring refinement network). We also show that WienerNet remains within 0.2 dB, on average, of the VRT transformer, outperforming it for noise STDs of $\sigma$ = 40 and $\sigma$ = 50, while being over ten times faster on the same hardware.

As described in Section~\ref{subsec:blind_denosier}, we also train and evaluate a single blind denoiser, WienerNet Blind, which requires no noise input, instead generating its own noise map. We show that with no user noise input, this network outperforms our optimised Wiener filter by \textbf{+ 3.1 dB} (+ 0.1231 SSIM) on average, remaining within 0.4 dB of our non-blind denoiser, with very few extra parameters and almost identical run times. Like the non-blind version, this denoiser also outperforms the VRT transformer at high noise levels.

Lastly, we evaluate the blind denoiser using a multi-scale denoising approach, by denoising each sequence at block sizes of 16, 32 and 64, and averaging the output, as described in Section~\ref{sec:mocomp_and_multiscale}. This method improves the performance of the blind denoiser at $\sigma$ = 10 and $\sigma$ = 20. This suggests an optimal weighted average exists, per noise level, which outperforms the single-scale approach.
\input{final_table}

\subsection{Motion Compensation}
Lastly, we evaluate motion compensation as a preprocessing step to denoising for both trained and untrained (WienerNetBlind) filters using Deepflow~\cite{weinzaepfel2013deepflow} and RAFT~\cite{teed2020raft} optical flow algorithms. This is implemented in the same manner as DVDNet~\cite{tassano2019dvdnet}.

In the untrained case, DeepFlow and RAFT do not improve denoising results in terms of PSNR but improve SSIM at all noise levels except for $\sigma$=10. This result may be attributed to occlusions created in the motion-compensated frames. 

For the trained case, Deepflow matches the non-motion compensated network at $\sigma$ = 10 and outperforms it at $\sigma$ = 20 and $\sigma$ = 30. However, SSIM results do not improve when motion compensation is applied to the trained network and no PSNR improvements are made at $\sigma$ = 40 and $\sigma$ = 50. These results may indicate some denoisers have been trained to handle the occlusions generated by motion compensation algorithms. In our case, more performance may be extracted if we discard or ignore frames which exceed a threshold value for occluded pixels.   
\input{images}

\begin{table}[t]
\centering
\resizebox{0.75\columnwidth}{!}{%
\begin{tabular}{@{}lllll@{}}
\cmidrule(r){1-4}
 &
  \multicolumn{1}{c}{None} &
  \multicolumn{1}{c}{Deepflow} &
  \multicolumn{1}{c}{Raft} &
   \\ 
Sigma &
  \multicolumn{1}{c}{PSNR / SSIM} &
  \multicolumn{1}{c}{PSNR / SSIM} &
  \multicolumn{1}{c}{PSNR / SSIM} &
   \\ \cmidrule(r){1-4}
10 & \textbf{35.67} / \textbf{0.9106} & 34.87 / 0.9071 &  34.94 / 0.9074  & \multirow{5}{*}{\rotatebox{90}{Wiener Opt.}} \\
20 & \textbf{31.84} / 0.8441 & 31.51 / 0.8480 &  31.68 / \textbf{0.8502}  &                          \\
30 & \textbf{29.74} / 0.7925 & 29.49 / 0.8016 &  29.53 / \textbf{0.8022}  &                            \\
40 & \textbf{27.97} / 0.7466 & 27.81 / \textbf{0.7591} &  27.75 / 0.7567  &                            \\
50 & \textbf{26.42} / 0.7063 & 26.30 / \textbf{0.7196} &  26.17 / 0.7137  &                            \\ \hdashline
10 & \textbf{36.38} / \textbf{0.9480} &  \textbf{36.38} / 0.9337 &  36.35 / 0.9338 & \multirow{5}{*}{\rotatebox{90}{WienerNetB.}}   \\
20 & 33.14 / \textbf{0.9071} &  \textbf{33.74} / 0.8914 &  33.62 / 0.8903 &                            \\
30 & 31.92 / \textbf{0.8823} & \textbf{31.99} / 0.8593 & 31.73 / 0.8556 &                            \\
40 & \textbf{30.69} / \textbf{0.8534} & 30.65 / 0.8310 &  30.24 / 0.8232 &                            \\
50 & \textbf{29.71} / \textbf{0.8261} &  29.51 / 0.8044  & 28.96 / 0.7915 &                            \\ \cmidrule(r){1-4}

\end{tabular}%
}
\caption{Motion compensation efficacy before and after training the Wiener Refinement network. Evaluation carried out on all datasets, $\sigma$=[10-50] where WienerNetB represents our blind denoiser.}
\label{tab:mo_comp}
\end{table}

\section{Conclusions}
In our work, we have demonstrated the efficiency of using small ancillary CNNs to improve the performance of a classic, optimised Bayesian filter, moving away from the black-box approach of CNN and transformer-based denoisers. Our denoiser is smaller in terms of parameters than all tested networks, and faster than the most competitive methods. We have also shown that current motion compensation methods do not always improve denoising performance. This may be optimised in future work, along with further improvements in denoising speed and weighted averaging for multi-scale denoising.

%% file: final_table.tex
\begin{table*}[t]
\centering
\resizebox{\textwidth}{!}
{
\begin{tabular}{@{}cccccccccc@{}}
\toprule
\multicolumn{1}{l}{} &
  \multicolumn{1}{c}{Noisy} &
  \multicolumn{1}{c}{DVDNet~\cite{tassano2019dvdnet}} &
  \multicolumn{1}{c}{FastDVDNet~\cite{tassano2020fastdvdnet}} &
  \multicolumn{1}{c}{VRT\cite{liang2022vrt}} &
  \multicolumn{1}{c}{VNLB\cite{arias2018video}} &
  \multicolumn{1}{c}{Wiener Opt.} &
  \multicolumn{1}{c}{WienerNet} &
  \multicolumn{1}{c}{WienerNetBlind} &
  \multicolumn{1}{c}{WienerNetBlind+MS} \\ 
STD &
  \multicolumn{1}{c}{PSNR / SSIM} &
  \multicolumn{1}{c}{PSNR / SSIM} &
  \multicolumn{1}{c}{PSNR / SSIM} &
  \multicolumn{1}{c}{PSNR / SSIM} &
  \multicolumn{1}{c}{PSNR / SSIM} &
  \multicolumn{1}{c}{PSNR / SSIM} &
  \multicolumn{1}{c}{PSNR / SSIM} &
  \multicolumn{1}{c}{PSNR / SSIM} &
  \multicolumn{1}{c}{PSNR / SSIM} \\ \midrule
10 &
  28.37 / 0.6862 &
  35.41 / 0.9174 &
  35.48 / 0.9180 &
  \textbf{37.59} / 0.9307 &
  37.58 / 0.9257 &
  34.54 / 0.8941 &
  37.14 / \textbf{0.9538} &
  36.38 / 0.9480 &
  36.41 / 0.9469 \\
20 &
  22.52 / 0.4431 &
  32.75 / 0.8724 &
  32.72 / 0.8700 &
  \textbf{34.55} / 0.8957 &
  33.82 / 0.8720 &
  30.82 / 0.8129 &
  33.78 / \textbf{0.9148} &
  33.14 / 0.9071 &
  33.54 / 0.9112 \\
30 &
  19.22 / 0.3119 &
  30.58 / 0.8292 &
  30.67 / 0.8277 &
  \textbf{32.28} / 0.8661 &
  31.13 / 0.8178 &
  28.65 / 0.7480 &
  31.96 / 0.8817 &
  31.92 / \textbf{0.8823} &
  31.80 / 0.8794 \\
40 &
  16.97 / 0.2325 &
  28.66 / 0.7818 &
  28.84 / 0.7863 &
  30.22 / 0.8354 &
  28.90 / 0.7665 &
  26.99 / 0.6950 &
  \textbf{30.71} / \textbf{0.8539} &
  30.69 / 0.8534 &
  30.56 / 0.8503 \\
50 &
  15.27 / 0.1804 &
  26.87 / 0.7345 &
  27.14 / 0.7461 &
  28.29 / 0.8029 &
  26.96 / 0.7195 &
  25.56 / 0.6515 &
  \textbf{30.35} / \textbf{0.8528} &
  29.71 / 0.8261 &
  29.59 / 0.8232 \\ \midrule
\multicolumn{1}{l}{Time (s)} &
 - &
 4.2k  &
 70.15  &
 1.9k  &
 26.6k &
  23.30 &
  149.26 & 
  149.92 &  
  2.0k \\
\multicolumn{1}{l}{Params (M)} &
  - &
  1.33 &
  2.50 &
  35.60 &
  - &
  - &
  0.29 & 
  0.29 &
  0.86 \\ \bottomrule
\end{tabular}
}
\caption{Quality benchmark of popular denoisers compared to WienerNet in PSNR (dB) and SSIM ([0-1]). Time is the total time taken to denoise the 10 test sequences, in seconds. Params is the number of trainable parameters in the deep denoisers. \textit{Wiener Opt.} is our denoiser without the coring refinement network, \textit{WienerNet} is our non-blind denoiser with the coring refinement network, \textit{WienerNetBlind} is our blind denoiser and \textit{WienerNetBlind+MS} is our multiscale blind denoiser.}
\label{tab:summary_table}
\end{table*}

%% file: images.tex
\begin{figure*}[t]
    \centering 
\begin{subfigure}{0.24\linewidth}
  \begin{tikzpicture}[spy using outlines={red,magnification=2.5,size=1.5cm, connect spies}]
\node  {\includegraphics[width=\linewidth]{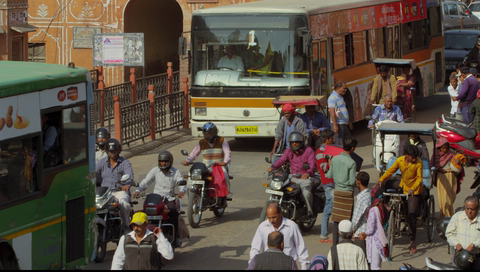}};
\spy on (1,-0.5) in node [left] at (-0.6,0.4);
\end{tikzpicture}
  \caption{Clean}
  \label{fig:1}
\end{subfigure}\hfil 
\begin{subfigure}{0.24\textwidth}
  \begin{tikzpicture}[spy using outlines={red,magnification=2.5,size=1.5cm, connect spies}]
\node  {\includegraphics[width=\linewidth]{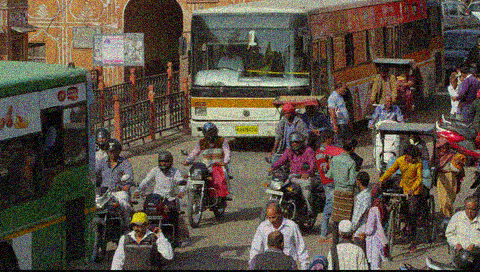}};
\spy on (1,-0.5) in node [left] at (-0.6,0.4);
\end{tikzpicture}
  \caption{Noisy}
  \label{fig:2}
\end{subfigure}\hfil 
\begin{subfigure}{0.24\textwidth}
  \begin{tikzpicture}[spy using outlines={red,magnification=2.5,size=1.5cm, connect spies}]
\node  {\includegraphics[width=\linewidth]{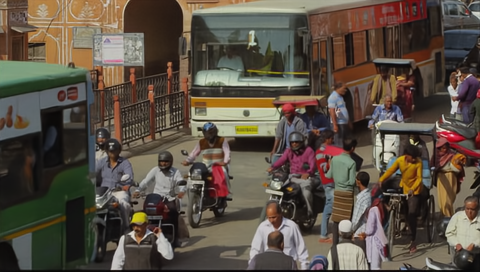}};
\spy on (1,-0.5) in node [left] at (-0.6,0.4);
\end{tikzpicture}
  \caption{DVDNet}
  \label{fig:3}
\end{subfigure}\hfil
\begin{subfigure}{0.24\textwidth}
\begin{tikzpicture}[spy using outlines={red,magnification=2.5,size=1.5cm, connect spies}]
\node  {\includegraphics[width=\linewidth]{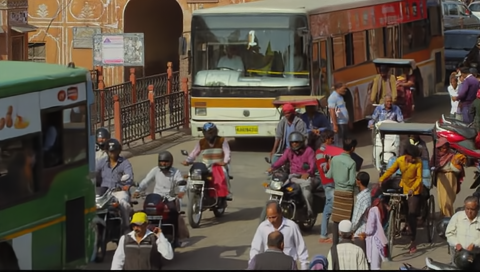}};
\spy on (1,-0.5) in node [left] at (-0.6,0.4);
\end{tikzpicture}
  \caption{FastDVDNet}
  \label{fig:4}
\end{subfigure}

\smallskip
\begin{subfigure}{0.24\textwidth}
  \begin{tikzpicture}[spy using outlines={red,magnification=2.5,size=1.5cm, connect spies}]
\node  {\includegraphics[width=\linewidth]{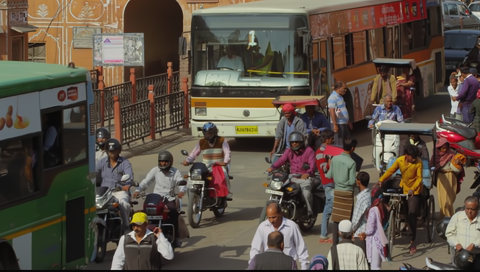}};
\spy on (1,-0.5) in node [left] at (-0.6,0.4);
\end{tikzpicture}
  \caption{VRT}
  \label{fig:5}
\end{subfigure}\hfil 
\begin{subfigure}{0.24\textwidth}
  \begin{tikzpicture}[spy using outlines={red,magnification=2.5,size=1.5cm, connect spies}]
\node  {\includegraphics[width=\linewidth]{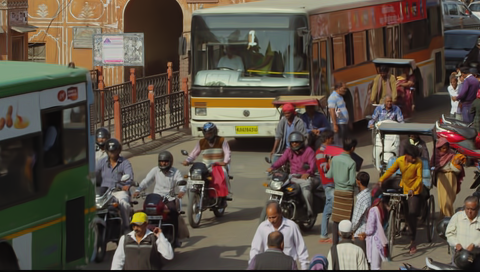}};
\spy on (1,-0.5) in node [left] at (-0.6,0.4);
\end{tikzpicture}
  \caption{VNLB}
  \label{fig:6}
\end{subfigure}\hfil 
\begin{subfigure}{0.24\textwidth}
  \begin{tikzpicture}[spy using outlines={red,magnification=2.5,size=1.5cm, connect spies}]
\node  {\includegraphics[width=\linewidth]{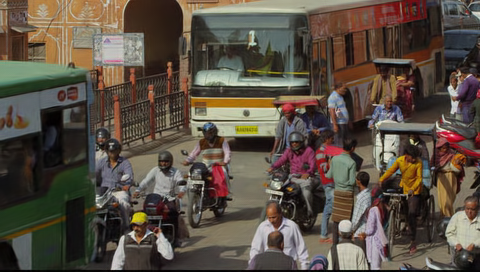}};
\spy on (1,-0.5) in node [left] at (-0.6,0.4);
\end{tikzpicture}
  \caption{WienerNet Non-Blind}
  \label{fig:7}
\end{subfigure}\hfil
\begin{subfigure}{0.24\textwidth}
  \begin{tikzpicture}[spy using outlines={red,magnification=2.5,size=1.5cm, connect spies}]
\node  {\includegraphics[width=\linewidth]{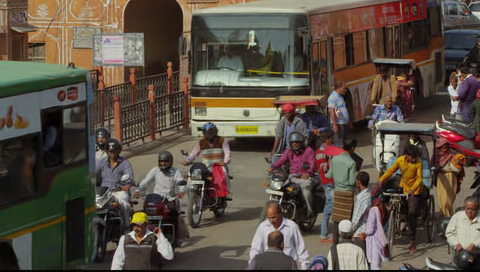}};
\spy on (1,-0.5) in node [left] at (-0.6,0.4);
\end{tikzpicture}
  \caption{WienerNet Blind}
  \label{fig:8}
\end{subfigure}
\caption{Sample output frame at $\sigma$ = 20, taken from benchmark scenes. For complete sequences, please visit our .\href{https://github.com/MrBled/WienerNet-ICME}{Supplementary Material Repository.}}
\label{fig:images20}
\end{figure*}

%% file: conference_101719.bbl
\begin{thebibliography}{10}
\providecommand{\url}[1]{#1}
\csname url@samestyle\endcsname
\providecommand{\newblock}{\relax}
\providecommand{\bibinfo}[2]{#2}
\providecommand{\BIBentrySTDinterwordspacing}{\spaceskip=0pt\relax}
\providecommand{\BIBentryALTinterwordstretchfactor}{4}
\providecommand{\BIBentryALTinterwordspacing}{\spaceskip=\fontdimen2\font plus
\BIBentryALTinterwordstretchfactor\fontdimen3\font minus \fontdimen4\font\relax}
\providecommand{\BIBforeignlanguage}[2]{{%
\expandafter\ifx\csname l@#1\endcsname\relax
\typeout{** WARNING: IEEEtran.bst: No hyphenation pattern has been}%
\typeout{** loaded for the language `#1'. Using the pattern for}%
\typeout{** the default language instead.}%
\else
\language=\csname l@#1\endcsname
\fi
#2}}
\providecommand{\BIBdecl}{\relax}
\BIBdecl

\bibitem{pratt1972generalized}
W.~K. Pratt, ``Generalized wiener filtering computation techniques,'' \emph{IEEE Transactions on Computers}, vol. 100, no.~7, pp. 636--641, 1972.

\bibitem{king1983wiener}
M.~A. King, P.~W. Doherty, R.~B. Schwinger, and B.~C. Penney, ``A wiener filter for nuclear medicine images,'' \emph{Medical physics}, vol.~10, no.~6, pp. 876--880, 1983.

\bibitem{giger1984investigation}
M.~L. Giger, K.~Doi, and C.~E. Metz, ``Investigation of basic imaging properties in digital radiography. 2. noise wiener spectrum,'' \emph{Medical physics}, vol.~11, no.~6, pp. 797--805, 1984.

\bibitem{benesty2010study}
J.~Benesty, J.~Chen, and Y.~Huang, ``Study of the widely linear wiener filter for noise reduction,'' in \emph{2010 IEEE International Conference on Acoustics, Speech and Signal Processing}.\hskip 1em plus 0.5em minus 0.4em\relax IEEE, 2010, pp. 205--208.

\bibitem{mallat1989theory}
S.~G. Mallat, ``A theory for multiresolution signal decomposition: the wavelet representation,'' \emph{IEEE transactions on pattern analysis and machine intelligence}, vol.~11, no.~7, pp. 674--693, 1989.

\bibitem{combettes2004wavelet}
P.~L. Combettes and J.-C. Pesquet, ``Wavelet-constrained image restoration,'' \emph{International Journal of Wavelets, Multiresolution and Information Processing}, vol.~2, no.~04, pp. 371--389, 2004.

\bibitem{malfait1997wavelet}
M.~Malfait and D.~Roose, ``Wavelet-based image denoising using a markov random field a priori model,'' \emph{IEEE Transactions on image processing}, vol.~6, no.~4, pp. 549--565, 1997.

\bibitem{dabov2007image}
K.~Dabov, A.~Foi, V.~Katkovnik, and K.~Egiazarian, ``Image denoising by sparse 3-d transform-domain collaborative filtering,'' \emph{IEEE Transactions on image processing}, vol.~16, no.~8, pp. 2080--2095, 2007.

\bibitem{maggioni2012video}
M.~Maggioni, G.~Boracchi, A.~Foi, and K.~Egiazarian, ``Video denoising, deblocking, and enhancement through separable 4-d nonlocal spatiotemporal transforms,'' \emph{IEEE Transactions on image processing}, vol.~21, no.~9, pp. 3952--3966, 2012.

\bibitem{arias2018video}
P.~Arias and J.-M. Morel, ``Video denoising via empirical bayesian estimation of space-time patches,'' \emph{Journal of Mathematical Imaging and Vision}, vol.~60, no.~1, pp. 70--93, 2018.

\bibitem{8803314}
A.~Davy, T.~Ehret, J.-M. Morel, P.~Arias, and G.~Facciolo, ``A non-local cnn for video denoising,'' in \emph{2019 IEEE International Conference on Image Processing (ICIP)}, 2019, pp. 2409--2413.

\bibitem{tassano2019dvdnet}
M.~Tassano, J.~Delon, and T.~Veit, ``Dvdnet: A fast network for deep video denoising,'' in \emph{2019 IEEE International Conference on Image Processing (ICIP)}.\hskip 1em plus 0.5em minus 0.4em\relax IEEE, 2019, pp. 1805--1809.

\bibitem{tassano2020fastdvdnet}
------, ``Fastdvdnet: Towards real-time deep video denoising without flow estimation,'' in \emph{Proceedings of the IEEE/CVF conference on computer vision and pattern recognition}, 2020, pp. 1354--1363.

\bibitem{claus2019videnn}
M.~Claus and J.~Van~Gemert, ``Videnn: Deep blind video denoising,'' in \emph{Proceedings of the IEEE/CVF conference on computer vision and pattern recognition workshops}, 2019, pp. 0--0.

\bibitem{yue2020supervised}
H.~Yue, C.~Cao, L.~Liao, R.~Chu, and J.~Yang, ``Supervised raw video denoising with a benchmark dataset on dynamic scenes,'' in \emph{Proceedings of the IEEE/CVF conference on computer vision and pattern recognition}, 2020, pp. 2301--2310.

\bibitem{vaksman2021patch}
G.~Vaksman, M.~Elad, and P.~Milanfar, ``Patch craft: Video denoising by deep modeling and patch matching,'' in \emph{Proceedings of the IEEE/CVF International Conference on Computer Vision}, 2021, pp. 2157--2166.

\bibitem{liang2022vrt}
J.~Liang, J.~Cao, Y.~Fan, K.~Zhang, R.~Ranjan, Y.~Li, R.~Timofte, and L.~Van~Gool, ``Vrt: A video restoration transformer,'' \emph{arXiv preprint arXiv:2201.12288}, 2022.

\bibitem{yue2023rvideformer}
H.~Yue, C.~Cao, L.~Liao, and J.~Yang, ``Rvideformer: Efficient raw video denoising transformer with a larger benchmark dataset,'' \emph{arXiv e-prints}, pp. arXiv--2305, 2023.

\bibitem{wang2019edvr}
X.~Wang, K.~C. Chan, K.~Yu, C.~Dong, and C.~Change~Loy, ``Edvr: Video restoration with enhanced deformable convolutional networks,'' in \emph{Proceedings of the IEEE/CVF Conference on Computer Vision and Pattern Recognition Workshops}, 2019, pp. 0--0.

\bibitem{liang2021swinir}
J.~Liang, J.~Cao, G.~Sun, K.~Zhang, L.~Van~Gool, and R.~Timofte, ``Swinir: Image restoration using swin transformer,'' in \emph{Proceedings of the IEEE/CVF International Conference on Computer Vision}, 2021, pp. 1833--1844.

\bibitem{zamir2022restormer}
S.~W. Zamir, A.~Arora, S.~Khan, M.~Hayat, F.~S. Khan, and M.-H. Yang, ``Restormer: Efficient transformer for high-resolution image restoration,'' in \emph{Proceedings of the IEEE/CVF Conference on Computer Vision and Pattern Recognition}, 2022, pp. 5728--5739.

\bibitem{10222826}
C.~Bled and F.~Pitié, ``Pushing the limits of the wiener filter in image denoising,'' in \emph{2023 IEEE International Conference on Image Processing (ICIP)}, 2023, pp. 2590--2594.

\bibitem{zhang2020deep}
K.~Zhang, L.~V. Gool, and R.~Timofte, ``Deep unfolding network for image super-resolution,'' in \emph{Proceedings of the IEEE/CVF conference on computer vision and pattern recognition}, 2020, pp. 3217--3226.

\bibitem{wiener1949extrapolation}
N.~Wiener, N.~Wiener, C.~Mathematician, N.~Wiener, N.~Wiener, and C.~Math{\'e}maticien, \emph{Extrapolation, interpolation, and smoothing of stationary time series: with engineering applications}.\hskip 1em plus 0.5em minus 0.4em\relax MIT press Cambridge, MA, 1949, vol. 113, no.~21.

\bibitem{kokaram19943d}
A.~Kokaram, ``3d wiener filtering for noise suppression in motion picture sequences using overlapped processing,'' \emph{Signal Processing VII, Vol3}, pp. 1780--1783, 1994.

\bibitem{zhang2017beyond}
K.~Zhang, W.~Zuo, Y.~Chen, D.~Meng, and L.~Zhang, ``Beyond a gaussian denoiser: Residual learning of deep cnn for image denoising,'' \emph{IEEE transactions on image processing}, vol.~26, no.~7, pp. 3142--3155, 2017.

\bibitem{mohan2020robust}
S.~Mohan, Z.~Kadkhodaie, E.~P. Simoncelli, and C.~Fernandez-Granda, ``Robust and interpretable blind image denoising via bias-free convolutional neural networks,'' in \emph{International Conference on Learning Representations}, 2020.

\bibitem{montgomery1994xiph}
C.~Montgomery and H.~Lars, ``Xiph. org video test media (derf’s collection),'' \emph{Online, https://media. xiph. org/video/derf}, vol.~6, 1994.

\bibitem{ma2021bvi}
D.~Ma, F.~Zhang, and D.~R. Bull, ``Bvi-dvc: A training database for deep video compression,'' \emph{IEEE Transactions on Multimedia}, vol.~24, pp. 3847--3858, 2021.

\bibitem{katsenou2020bvi}
A.~V. Katsenou, G.~Dimitrov, D.~Ma, and D.~R. Bull, ``Bvi-syntex: A synthetic video texture dataset for video compression and quality assessment,'' \emph{IEEE Transactions on Multimedia}, vol.~23, pp. 26--38, 2020.

\bibitem{loshchilov2018fixing}
I.~Loshchilov and F.~Hutter, ``Fixing weight decay regularization in adam,'' 2018.

\bibitem{weinzaepfel2013deepflow}
P.~Weinzaepfel, J.~Revaud, Z.~Harchaoui, and C.~Schmid, ``Deepflow: Large displacement optical flow with deep matching,'' in \emph{Proceedings of the IEEE international conference on computer vision}, 2013, pp. 1385--1392.

\bibitem{teed2020raft}
Z.~Teed and J.~Deng, ``Raft: Recurrent all-pairs field transforms for optical flow,'' in \emph{Computer Vision--ECCV 2020: 16th European Conference, Glasgow, UK, August 23--28, 2020, Proceedings, Part II 16}.\hskip 1em plus 0.5em minus 0.4em\relax Springer, 2020, pp. 402--419.

\end{thebibliography}
